\documentclass[prd,superscriptaddress,twocolumn,
floatfix,preprintnumbers,altaffilletter]{revtex4}
\usepackage{graphicx,url,amssymb,amsmath,longtable,rotating,color,units,
wasysym,subfigure,epsfig,multirow}

\begin{document}

\title{All-sky, narrowband, gravitational-wave radiometry with folded data}

\author{Eric~Thrane}
\email{eric.thrane@monash.edu}
\affiliation{School of Physics and Astronomy, Monash University, Clayton, Victoria 3800, Australia}

\author{Sanjit Mitra}
\affiliation{Pune University Campus, Pune 411007, India}

\author{Nelson~Christensen}
\affiliation{Physics and Astronomy, Carleton College, Northfield, Minnesota 55057, USA}

\author{Vuk~Mandic}
\affiliation{School of Physics and Astronomy, University of Minnesota, Minneapolis, Minnesota 55455, USA}

\author{Anirban Ain}
\affiliation{Pune University Campus, Pune 411007, India}

\begin{abstract}
  Gravitational-wave radiometry is a powerful tool by which weak signals with unknown signal morphologies are recovered through a process of cross correlation.
  Radiometry has been used, e.g., to search for persistent signals from known neutron stars such as Scorpius X-1.
  In this paper, we demonstrate how a more ambitious search---for persistent signals from {\em unknown} neutron stars---can be efficiently carried out using folded data, in which an entire $\sim$year-long observing run is represented as a single sidereal day.
  The all-sky, narrowband radiometer search described here will provide a computationally tractable means to uncover gravitational-wave signals from unknown, nearby neutron stars in binary systems, which can have modulation depths of $\approx$$0.1$--$\unit[2]{Hz}$.
  It will simultaneously provide a sensitive search algorithm for other persistent, narrowband signals from unexpected sources.
\end{abstract}

\maketitle

\section{Introduction}
Gravitational-wave radiometry is a technique by which two or more detectors are cross-correlated in order to identify signals in the form of excess coherence~\cite{radiometer,radio_method}.
It is especially well-suited to situations where it is either impossible or impractical to carry out a matched filter search due to theoretical uncertainty and/or the vastness of the signal parameter space.
LIGO~\cite{aligo} and Virgo~\cite{virgo} radiometer searches for persistent gravitational waves have yielded limits on the gravitational-wave strain from targets including Scorpius X-1, the Galactic Center, and Supernova 1987A~\cite{sph_results}.
Radiometer searches are also sensitive to hot spots created from the superposition of many persistent sources~\cite{Dhurandhar,Mazumder,ns_sgwb}.
By restricting the timescale of integration, the same method has been applied to search for relatively long-lived $\sim$$10$--$\unit[1000]{s}$ gravitational-wave transients~\cite{stamp}, e.g., associated with long gamma-ray bursts~\cite{lgrb}.
Radiometry has also been proposed in the context of gravitational-wave astronomy with pulsar timing arrays~\cite{anholm}.

Due to its robustness and efficiency, there is strong motivation to extend the parameter space of radiometer searches.
In this paper, we propose a method to extend the search for persistent narrowband signals to target all frequencies {\em and} all directions on the sky---not just ones associated with targets such as Scorpius X-1.
An efficient all-sky search for persistent narrowband gravitational waves could facilitate the detection, e.g., of electromagnetically quiet spinning neutron stars in binary systems.
The challenge is to overcome computational challenges that make the search difficult.

There is a significant burden from both data storage and computation time.
In previous {\em broadband} radiometer searches for persistent gravitational waves~\cite{sph_results}, a Fisher matrix was used to characterize the covariance between different patches of sky~\cite{sph}.
However, saving Fisher matrices for many frequency bins (and for many ``jobs'' associated with stretches of science quality data) creates a burdensome data storage problem.
The computational cost of analyzing a full year of data to look at thousands of frequency bins is both inefficient and potentially prohibitive in terms of computation time.

Recent work~\cite{folding} shows that by ``folding'' data into a single sidereal day, it is possible to represent an entire year of radiometer data with just one twenty-four long spectrogram.
We show that this several-hundred-fold reduction in data volume facilitates efficient searches for persistent narrowband signals.

The remainder of this paper is organized as follows.
In Section~\ref{motivation}, we discuss the motivation for narrowband all-sky searches.
Section~\ref{method} describes how previous work on folded data~\cite{folding} and searches for $\approx$day-long signals~\cite{verylong} can be combined to design an efficient all-sky narrowband radiometer.
In Section~\ref{demonstration}, we demonstrate an all-sky narrowband radiometer search using folded data.
Finally, in Section~\ref{conclusions}, we discuss the astrophysical implications of these results.

\section{Motivation}\label{motivation}
One of the primary targets for an all-sky narrowband search for persistent gravitational waves is spinning neutron stars in binary systems.
The motivation for such a search was recently laid out in~\cite{twospect}.
In a variety of scenarios, accretion from a companion star can induce a time-varying quadrupole moment, which, in some cases, may persist after accretion abates.
For example, persistent localized mass accumulation may occur due to magnetic fields~\cite{bildsten} depending on unknown details of neutron star physics~\cite{vigelius1,priymak,wette,vigelius2,vigelius3}.
Magnetic fields may also induce deformations in the stellar interior~\cite{melatos2}.
Rotational instabilities such as $r$-modes may be sustained through accretion~\cite{reisenegger,ushomirsky}.

When the neutron star in a binary is a pulsar, it is possible to carry out an optimal search using matched filtering.
However, if it is electromagnetically quiet, then the space of possible signals becomes prohibitively large for a fully coherent search.
When at least the sky location is known, one can apply a targeted radiometer search~\cite{radiometer,radio_method,sph_results} or other semicoherent methods~\cite{twospect,twospect_method,sideband_method,crosscorr,polynomial}.
If the sky location is unknown, the problem is significantly more challenging and the signal might be missed or vetoed as a noise artifact by non-specialized pipelines~\cite{twospect}.

The search technique proposed here is highly complementary to the semicoherent method described in~\cite{twospect,twospect_method}.
The ``TwoSpect'' algorithm~\cite{twospect,twospect_method} employs a canonical signal model in which a neutron star emits periodic waves, modulated by binary motion.
The radiometer, on the other hand, employs only minimal assumptions; namely, that the signal is persistent and narrowband.
By incorporating additional assumptions about the signal model, a tuned search gains sensitivity.
The advantage of the radiometer, in contrast, is that it is extremely robust since it relies only on excess coherence in two or more detectors.
Indeed, while we focus on spinning neutron stars in binary systems, the narrowband radiometer is sensitive to {\em any} persistent narrowband source, whether or not it is modulated by binary motion.

\section{Method}\label{method}
\subsection{Folding}
We utilize the idea of folded data as described in~\cite{folding}, which we reformulate in the language of recent radiometer developments~\cite{stamp,stochtrack,lgrb,stochsky,stochtrack_cbc,stochtrack_ecbc,verylong}.
During an observing run, typically lasting a few months to a few years, data is collected from two spatially separated detectors.
(This discussion straightforwardly extends to three or more detectors, but our presentation focuses on the two-detector case for simplicity.)
We parse the data for which both detectors are simultaneously operational into sidereal days ($\unit[23]{hr}$, $\unit[56]{min}$, $\unit[4]{s}$).
Every sidereal day of radiometer data can be represented using a complex-valued estimator~\cite{verylong}:
\begin{equation}\label{eq:Y}
    \widehat{\mathfrak{Y}}(t;f) \equiv \frac{2}{\cal N} 
    \tilde{s}_1^*(t;f) \tilde{s}_2(t;f)
\end{equation}
\begin{equation}\label{eq:sigma}
    \sigma_{\mathfrak{Y}}(t;f) \equiv \frac{1}{2} \sqrt{P'_1(t;f) P'_2(t;f)}.
\end{equation}
Here, $(t;f)$ are spectrogram indices: $t$ is the start time of each data segment and $f$ is the frequency of the Fourier transformed data associated with that segment.
Each $\tilde{s}_I(t;f)$ refers to the discrete Fourier transform of the strain data from detector $I$.
The variable ${\cal N}$ is a Fourier normalization constant, $\sigma_{\mathfrak{Y}}(t;f)$ is an estimator for the uncertainty associated with $ \widehat{\mathfrak{Y}}$, and each $P'_I(t;f)$ represents the strain auto-power measured in detector $I$.
(The prime denotes that the auto-power is calculated as an average of neighboring segments.)
The direction of the source is encoded in the phase of $\mathfrak{Y}(t;f)$, which depends on the time delay between the two detectors~\cite{stochsky}.

Since the signal-induced phase delay and detectors' antenna factors are periodic on the timescale of a sidereal day, it is possible to define folded estimators by summing data over many sidereal days:
\begin{equation}
    \widehat{\mathfrak{Y}}^\text{fold}(t;f) = \sum_k 
    \widehat{\mathfrak{Y}}(t;f|k) \sigma_{\mathfrak{Y}}^{-2}(t;f|k) \Big/
    \sum_k \sigma_{\mathfrak{Y}}^{-2}(t;f|k)
\end{equation}
\begin{equation}
    \sigma_{\mathfrak{Y}}^\text{fold}(t;f) = 
    \left( \sum_k \sigma_{\mathfrak{Y}}^{-2}(t;f|k) \right)^{-1/2} .
\end{equation}
Here, $k$ runs over all the sidereal days of the observing run.

Following~\cite{stochsky}, we define complex signal-to-noise ratio for folded data:
\begin{equation}\label{eq:rho_fold}
  \mathfrak{p}^\text{fold}(t;f) \equiv \widehat{\mathfrak{Y}}^\text{fold}(t;f)/
  \sigma_{\mathfrak{Y}}^\text{fold}(t;f) .
\end{equation}
The variable $\mathfrak{p}^\text{fold}(t;f)$ is a complex-valued spectrogram; see Fig.~\ref{fig:add_days}.
Gravitational-waves induce excess $\mathfrak{p}(t;f)$, which appears visually as brighter than usual spectrogram pixels.

By folding a year of data into a sidereal day, the volume of data is reduced by a factor of $\approx$$365$, which dramatically reduces both storage needs and computation time.
And, if we restrict our attention to persistent narrowband signals, then the folding operation is {\em lossless} in the sense that there is no more information (useful in a search for persistent signals) in a year of cross-correlated data than there is in a folded day.

\subsection{Radiometry}
By specializing the methods from~\cite{verylong} to consider approximately monochromatic signals lasting exactly one sidereal day, and then applying these methods to folded data~\cite{folding}, it is possible to construct a very efficient algorithm for the detection of electromagnetically quiet binary neutron stars (and other narrowband sources not matching a standard continuous wave model).
Note that when we refer in this paper to signals as ``approximately monochromatic,'' we mean that they fit within single $\approx$$\unit[1]{Hz}$ frequency bin.
Such signals include neutron stars in binary systems, which are relatively broadband compared to isolated neutron stars.
However, the signal is narrowband in the sense that it is contained within just one radiometer frequency bin.
In reality, the modulation depth of neutron stars in binaries can range from $\approx$$0.1$--$\unit[2]{Hz}$~\cite{twospect}, so a more nuanced definition of ``narrowband,'' employing variable frequency bin sizes, is appropriate.
However, this is beyond our present scope.

Radiometer searches can be cast as a pattern recognition problems in which an algorithm looks for statistically significant clusters of pixels~\cite{stamp}.
As we are focused here on approximately monochromatic signals, the optimal pattern recognition algorithm is a simple sum over the frequency bins of $\mathfrak{p}^\text{fold}$.
In the language of seedless clustering~\cite{stochtrack,stochsky}, each signal template corresponds to a different row in the $\mathfrak{p}^\text{fold}$ spectrogram.
We assume that the signal starts at the beginning of the spectrogram and finishes at the end.
To assume otherwise would imply an unphysical signal that starts and stops with a period matching Earth's sidereal day.
Likewise, it would not make sense to allow for any frequency evolution since this would imply that the signal repeats this evolution with the period of a sidereal day.

The following expression for the detection statistic $\text{SNR}_\text{tot}(f|\hat\Omega)$ is specialized from~\cite{verylong} to focus on monochromatic signals:
\begin{equation}\label{eq:vlong_as}
  \text{SNR}_\text{tot}(f|\hat\Omega) = \frac{\text{Re}\left[
    \sum_{t}
    e^{\left(2\pi i f \hat\Omega \cdot \Delta \vec{x}(t)/c\right)} 
    \mathfrak{p}^\text{fold}(t;f) \, \epsilon_{12}(t|\hat\Omega)
    \right]
  }{
    \Big(\sum_{t}\epsilon_{12}^2(t|\hat\Omega)\Big)^{1/2}
  } .
\end{equation}
Here, $\hat\Omega$ is the direction of the source, $\Delta \vec{x}$ is the difference in detector position, and $c$ is the speed of light.
The phase factor $e^{\left(2\pi i f \hat\Omega \cdot \Delta \vec{x}(t)/c\right)}$ ``points'' the spectrogram toward the source by rotating the phase angle of $\mathfrak{p}^\text{fold}$ to zero so that the observed signal-to-noise ratio is real and positive~\cite{stochsky,verylong}.
The variable $\epsilon_{12}(t|\hat\Omega)$ is a time-dependent efficiency factor characterizing the fraction of gravitational-wave power measured due to the non-unity antenna factors of gravitational-wave interferometers:
\begin{equation}\label{eq:epsilon}
  \epsilon_{12}(t|\hat\Omega) \equiv \frac{1}{2}
  \sum_A F_1^A(t|\hat\Omega) F_2^A(t|\hat\Omega) .
\end{equation}
Here, $F_I^A(t|\hat\Omega)$ is the antenna factor~\cite{300years} for detector $I$ and $A=+,\times$ are the different polarization states.
As the Earth rotates, the detectors become more/less favorably aligned relative to the source.
The sum over $t$ is carried out over all the (typically $\approx$$\unit[60]{s}$) segments in a sidereal day.
For additional information pertaining to Eq.~\ref{eq:vlong_as}, the interested reader is referred to~\cite{verylong} and reference therein.

For the sake of compact notation, Eq.~\ref{eq:vlong_as} assumes that detector noise is approximately constant over the span of a sidereal day.
However, we note that it is straightforward to generalize to the case of non-stationary noise by weighting each data segment appropriately using a standard inverse variance weighting method~\cite{stoch_allenromano}.

Previous work~\cite{stochsky,verylong,stochtrack_cbc,stochtrack_ecbc} has shown that Eq.~\ref{eq:rho_fold} and Eq.~\ref{eq:vlong_as} can serve as the basis for ``all-sky'' searches for which the signal time and direction are unknown a priori.
Many directions in the sky can be considered simultaneously using parallelized computer code run on suitable processors, e.g., GPUs and multi-core CPUs~\cite{stochsky}.
For each observation, we record $\text{SNR}_\text{tot}^\text{max}$---the maximal value of $\text{SNR}_\text{tot}$ for all the directions:
\begin{equation}\label{eq:max}
  \text{SNR}_\text{tot}^\text{max}(f) = \max_{\hat\Omega}
  \left[ \text{SNR}_\text{tot}(f|\hat\Omega) \right] .
\end{equation}

Eq.~\ref{eq:vlong_as} assumes a fairly constrained signal model: approximately monochromatic signals, which start at the beginning of and finish at the of a sidereal day.
Thus, it is possible to search every sky location (given some resolution) using a HEALPix~\cite{healpix} grid.
In this work, we consider a number of sky locations chosen so as to be spaced at angular separations less than the diffraction limited resolution of the detector network.
For the LIGO network,
\begin{equation}
  \theta \approx \frac{c}{f} \frac{1}{\left|\Delta\vec{x}\right|} 
  \approx \left(\frac{\unit[1000]{Hz}}{f}\right) 5^\circ .
\end{equation}
In the demonstration described below, we use HEALPix to efficiently sample the sky with 3072 tiles.

\subsection{Comparison with targeted searches}
In order to highlight the usefulness of this framework, it is useful to compare it to the procedure used for previous narrowband {\em targeted} radiometer searches~\cite{radiometer,sph_results}, applied, e.g., to a single direction such as Scorpius X-1.
In such searches, one analyzes data in $\approx$$500$--$\unit[10000]{s}$-long ``jobs,'' defined as data stretches during which coincident data is available for both detectors.
During initial LIGO's fifth science run, there were $\approx$$20000$ jobs.
For each job, one calculates cross-power (Eq.~\ref{eq:Y}) and auto-powers (Eq.~\ref{eq:sigma}) by optimally combining the results from each of many (typically) $\unit[60]{s}$ segments in each job.
Then, during post-processing, the results from each job are optimally combined to produce estimators analogous to $\widehat{\mathfrak{Y}}$ and $\sigma_\mathfrak{p}$, but obtained by integrating over the entire observing run.

This procedure is repeated separately for each sky location.
Using this scheme, each sky location requires the generation of tens of thousands of output files, and so there is significant overhead associated with the analysis of more than a few sky locations.
Moreover, covariances between different patches on the sky---characterized by a Fisher matrix---lead to subtleties when interpreting the significance of an outlier~\cite{sph_results}.
In order to understand these covariances in this framework, one must therefore produce also a Fisher matrix for every frequency bin.

In the new paradigm we propose here, the data are first combined into a single sidereal day, which can be used to look in every sky direction.
This data is small enough to be easily manipulated on a typical computer.
Once the data are loaded, we can search over every sky location all at once without having to read or write additional files, thereby eliminating a significant bottleneck.
Also, this method eliminates the need to store Fisher matrices.
This is because the covariances between different sky locations are automatically encoded into the factors of $e^{\left(2\pi i f \hat\Omega \cdot \Delta \vec{x}/c\right)}$ and $\epsilon(t|\hat\Omega)$.

Thus, one can use numerical simulations, in which $\mathfrak{p}$ is drawn from a Gaussian distribution, in order to determine the false-alarm probability of obtaining a given value of $\text{SNR}_\text{tot}^\text{max}$.
(The assumption of Gaussianity is justified by invoking the central limit theorem, and it has been born out in previous measurements~\cite{stoch-S5,sph_results}.)
An example search, analyzing $\unit[750]{Hz}$ of bandwidth with $\Delta f=\unit[1]{Hz}$ resolution (see Fig.~\ref{fig:add_days}), and scanning over 3072 Healpixels, can be carried out in $\lesssim\unit[100]{s}$ using a single eight-core CPU.
This computation time does not include the modest time required for folding the data.

\section{Demonstration}\label{demonstration}
Our goal now is to demonstrate the recovery of a persistent narrowband signal in folded data without a priori knowledge of the sky location or frequency of the signal.
To begin, we generate twenty days of simulated data for the LIGO Hanford and LIGO Livingston detectors operating at design sensitivity.
The data consists of Gaussian noise plus a simulated monochromatic signal with frequency $f=\unit[600]{Hz}$ and strain amplitude $h_0=1.5\times10^{-24}$, located at $(\text{ra},\text{dec})=(\unit[18.5]{hr},39^\circ)$.
The signal is circularly polarized.
For our purposes, a monochromatic signal is a reasonable proxy for a signal modulated by binary motion (discussed in Section~\ref{motivation}) so long as the binary modulation is $\lesssim\unit[1]{Hz}$.

We generate coarse-grained spectrograms with a resolution of $(\unit[79]{s},\unit[1]{Hz})$ (with non-overlapping segments).
This frequency bin width is relatively well-matched to the observed modulation depth of known neutron stars in binary systems, which can range from~$\approx$$0.1$--$\unit[2]{Hz}$~\cite{twospect}.
(A more systematic study should employ variable bin widths to fully cover parameter space.)
We study a band ranging from $500$--$\unit[1250]{Hz}$, where many neutron stars in binary systems are expected to rotate~\cite{chakrabarty}.
Next, the data are folded into one sidereal day following the procedure outlined in Section~\ref{method}.
As the data are folded, we analyze the integrated spectrogram with every accumulated day in order to study how the signal grows with time.
Following the procedure described in Eqs.~\ref{eq:vlong_as} and~\ref{eq:max}, we scan 3072 Healpixels and record the maximum signal-to-noise ratio at each frequency $\text{SNR}_\text{tot}^\text{max}(f)$.

\begin{figure*}[hbtp!]
    \subfigure[]{\psfig{file=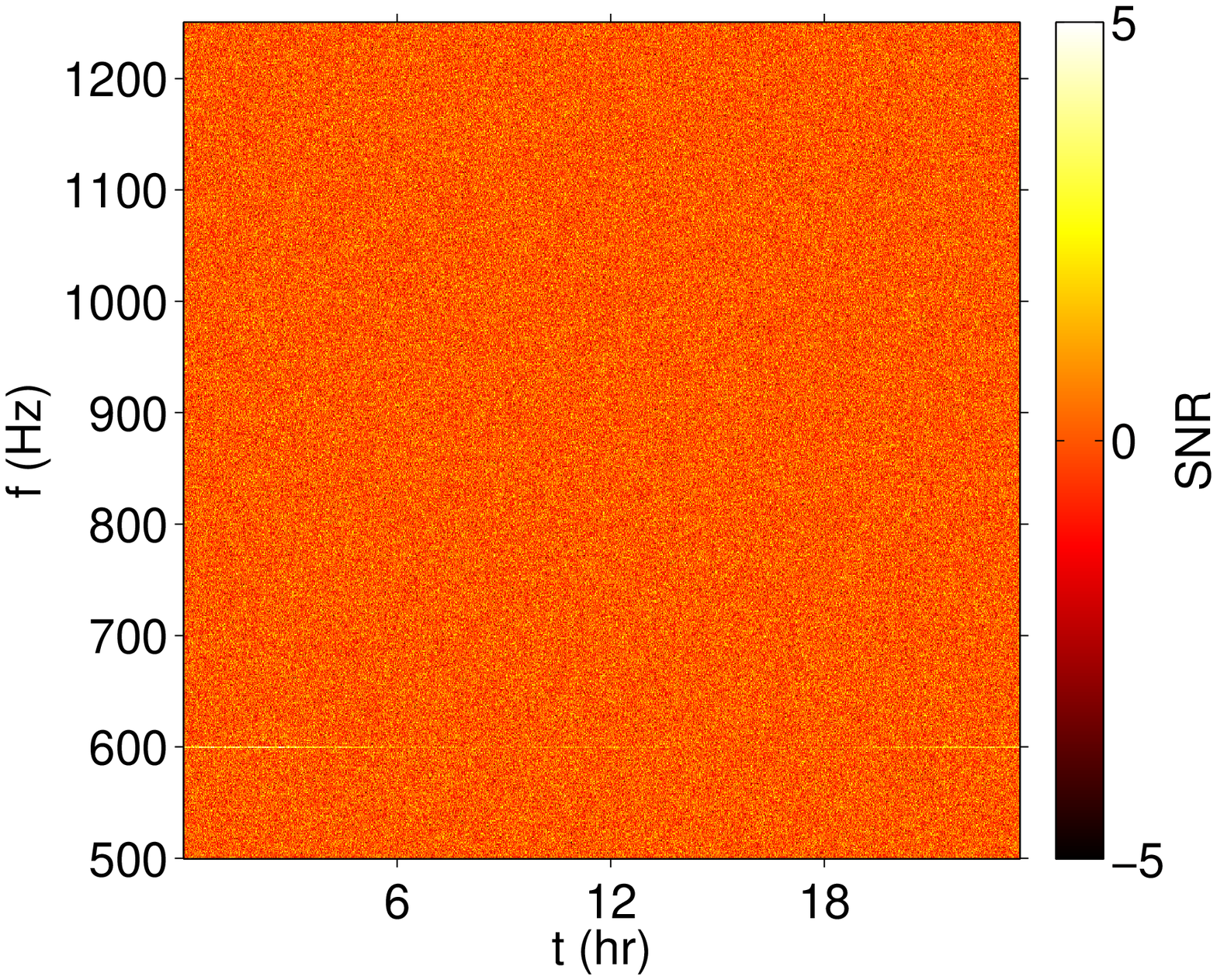, width=3in}} 
    \subfigure[]{\psfig{file=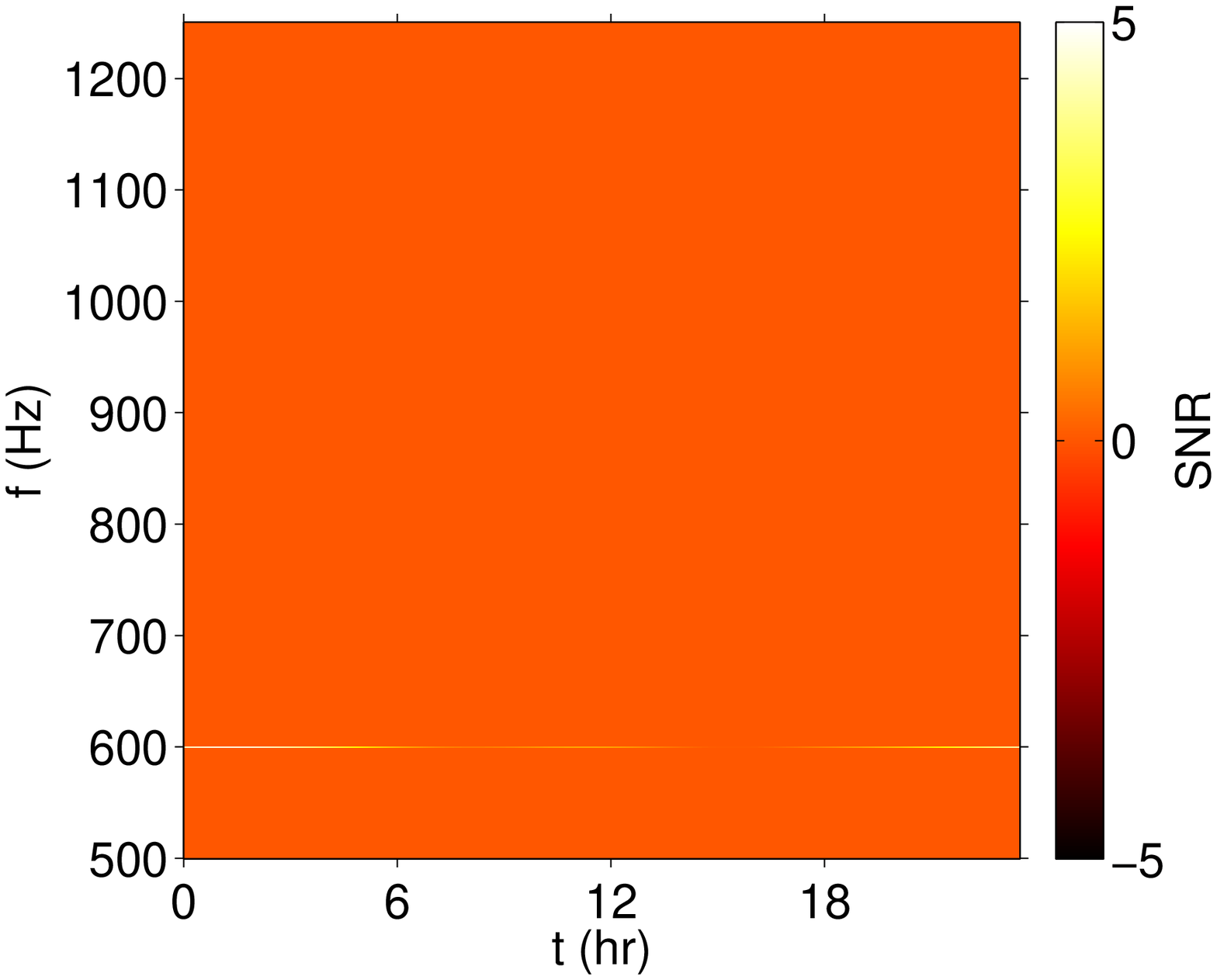, width=3in}}
    \subfigure[]{\psfig{file=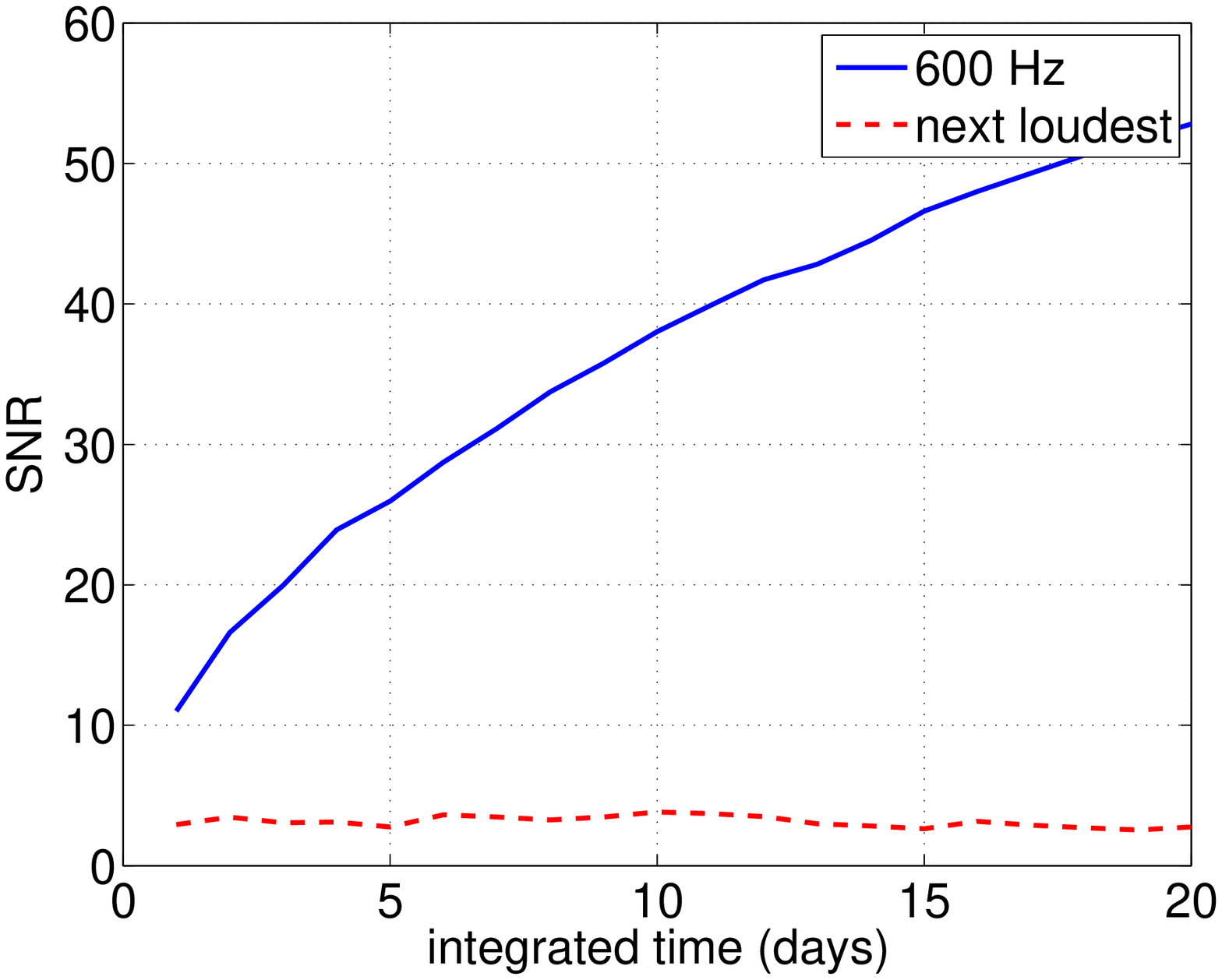, width=2.8in}}
    \subfigure[]{\psfig{file=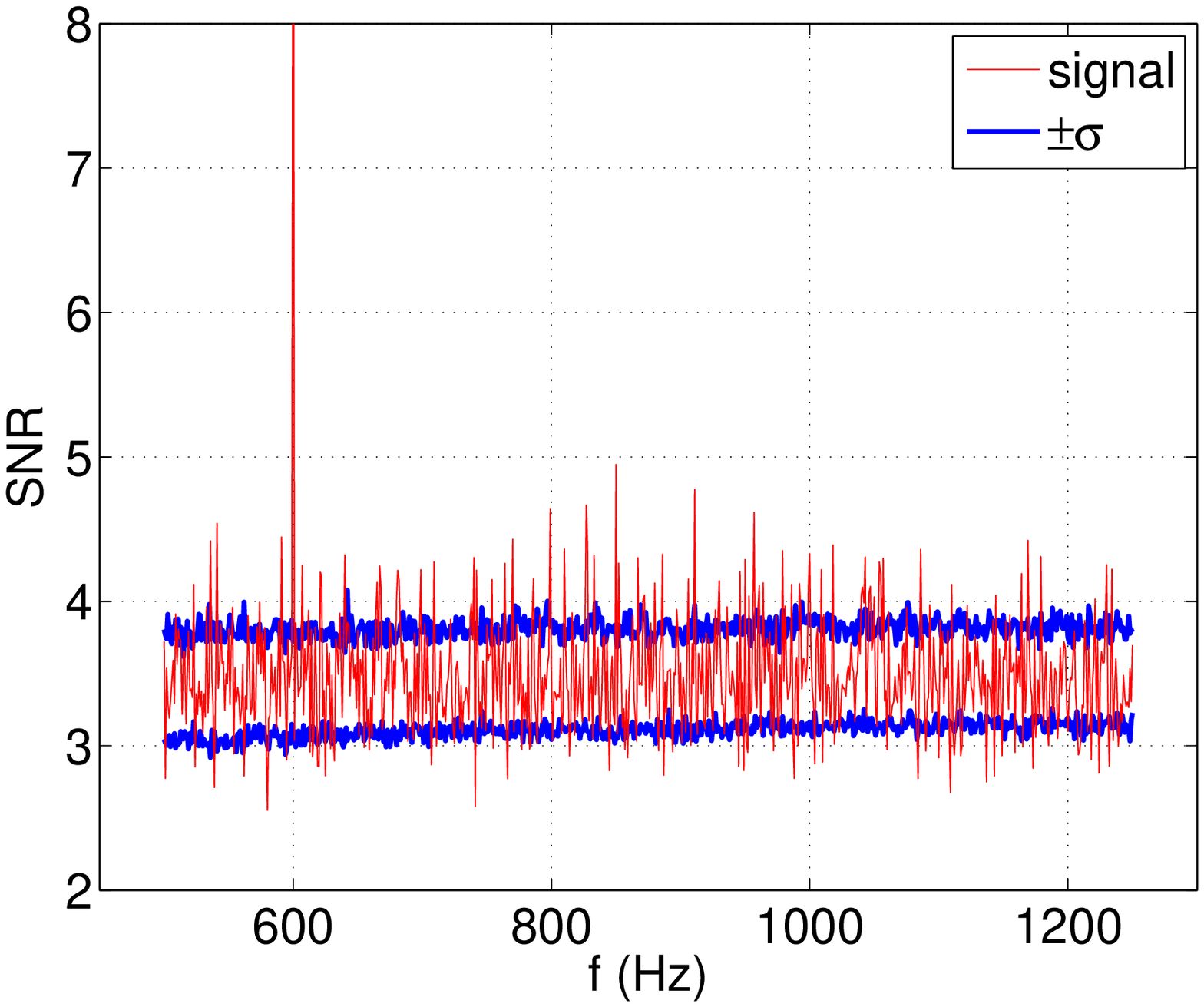, width=2.8in}}
 \caption{
   Narrowband signals in folded data.
   Top-left: a $(\delta t, \Delta f)=(\unit[79]{s},\unit[1]{Hz})$ spectrogram of the real part of $e^{\left(2\pi i f \hat\Omega \cdot \Delta \vec{x}/c\right)}\mathfrak{p}^\text{fold}(t;f)$ consisting of twenty days of data, folded into one sidereal day.
   The data consist of Advanced LIGO~\cite{aligo} Monte Carlo noise plus a simulated signal with amplitude $h_0=1.5\times10^{-24}$ and frequency $\unit[600]{Hz}$.
   The source is assumed to be face-on.
   The spectrogram is ``pointed'' in the correct direction using the appropriate phase factor to make the signal positive; see Eq.~\ref{eq:Y}.
   The simulated signal is detectable in just one day of data, but we show the result of 20 days of folding to make the signal visible by eye.
   The top-right panel shows the recovered signal in just one day of data.
   The signal is easily identified without any prior assumption about the source location or frequency.
   The bottom-left panel shows how the folded SNR grows with the addition of folded data.
   Blue shows the result for the bin with the signal while red is the next loudest bin.
   The bottom-right panel shows the maximum $\text{SNR}$ (scanning over the entire sky) as a function of frequency for one day of data (red).
   The injected signal is evident as a $\text{SNR}\approx12$ spike at $\unit[600]{Hz}$.
   Typical fluctuations due to pure noise are indicated with blue.
   It is interesting to note that there is a slight trend toward higher $\text{SNR}$ with increasing frequency due to increasing angular resolution.
   \label{fig:add_days}
 }
\end{figure*}

The results are summarized in Fig.~\ref{fig:add_days}.
In Fig.~\ref{fig:add_days}a, we show the real part of $e^{\left(2\pi i f \hat\Omega \cdot \Delta \vec{x}/c\right)}\mathfrak{p}^\text{fold}(t;f)$ representing twenty days of data folded into a single sidereal day.
Though the signal is virtually impossible to see with the naked eye in just one day of data, it is visible in this image at $\unit[600]{Hz}$ since the signal-to-noise ratio grows with continued integration.
Fig.~\ref{fig:add_days}b shows the recovered signal ($\text{SNR}_\text{tot}^\text{max}\approx12$) using just one day of data.
The excellent match in the time-dependent modulation suggest that the sky location is well matched.
Indeed, the best fit direction $(\text{ra},\text{dec})=(\unit[18.3]{hr},38.9^\circ)$ is the closest possible pixel to the true source location.

In Fig.~\ref{fig:add_days}c, we show how the signal-to-noise ratio of the $\unit[600]{Hz}$ signal (blue) grows as expected like the square root of the total observation time.
The dashed red curve shows the next-loudest frequency bin, which exhibits no tendency to grow or decline with continued integration.
In Fig.~\ref{fig:add_days}d, we show the signal-to-noise ratio spectrum for the spectrogram in Fig.~\ref{fig:add_days}a, maximized over search directions.
That is, for each frequency bin, we scanned the entire sky and plot (in red) $\text{SNR}_\text{tot}^\text{max}$ for the brightest patch of sky as a function of frequency.
The blue curves indicate typical one-sigma fluctuations for noise.
The $\unit[600]{Hz}$ signal is visible as a dramatic spike.
There is a slight tendency toward higher values of $\text{SNR}$ with increasing frequency as angular resolution improves and there are more independent patches of sky.

From Fig.~\ref{fig:add_days}, we show that it is possible carry out a computationally efficient search for persistent narrowband signals from all directions in the sky and at all frequencies.
Once the data have been folded, the entire search takes less than two minutes to carry out using a single eight-core CPU.

By studying the background distribution of $\text{SNR}_\text{tot}^\text{max}(f)$, we can estimate how the sensitivity of the radiometer search differs for a directed and an all-sky search.
Naively, we expect the all-sky sensitivity to be worse, but only slightly, since the additional trial factors incurred by looking in many directions will only slightly increase the detection threshold given the expected rapidly falling background distribution.
In order to identify a detection candidate with false alarm probability $1\%$, the targeted search requires $\text{SNR}_\text{tot}^\text{max}(f)\gtrsim4.5$ while the all-sky search requires $\text{SNR}_\text{tot}^\text{max}(f)\gtrsim5.8$ for the bandwidth and spectral resolution considered here.
This corresponds to a change in strain sensitivity of just $(5.8/4.5)^{1/2}-1\approx14\%$.

Upper limits can be set for each frequency bin independently, and so there are fewer trial factors to apply compared to the detection calculation, which must account for the existence of hundreds of frequency bins.
On average, the targeted search will set limits corresponding to $\text{SNR}_\text{tot}^\text{max}(f)\approx1.7$ whereas the all-sky search will set limits corresponding to $\text{SNR}_\text{tot}^\text{max}(f)\approx4$  (around $f\approx\unit[1000]{Hz}$ where we scan $\approx$$3072$ sky positions).
Thus, the all-sky limits are expected to be $\approx$$50\%$ higher in strain than limits obtained from a targeted search.
These estimates support the expectation that the all-sky search is only slightly less sensitive than the targeted search.

\section{Conclusions}\label{conclusions}
Rotating neutron stars in binaries are a very promising candidate for gravitational-wave detection by second-generation detectors like Advanced LIGO/Virgo.
Previous work~\cite{twospect_method} has enabled the exploration of these promising sources, though, no detection candidates have been found with initial LIGO data~\cite{twospect}.
Our presentation of an all-sky, narrowband radiometer using folded data is complementary since it makes only minimal assumptions about the source.
In this way, it should be sensitive not only to neutron stars in binary systems, but also to other sources, which emit approximately narrowband gravitational waves and which may or may not conform to the canonical model of an isolated neutron star.

A back-of-the-envelope calculation provides promising results.
In Subsection~\ref{demonstration}, we showed that a $h_0=1.5\times10^{-24}$ signal at $f=\unit[600]{Hz}$ can be easily recovered in one day of data.
Scaling to one year of data, this becomes $h_0\approx2\times10^{-25}$.
In terms of neutron-star ellipticity, this sensitivity can be restated~\cite{known_pulsars} as
\begin{equation}
  \begin{split}
  \epsilon \approx  1\times10^{-5} \left(\frac{0.4}{\beta}\right)
  \left(\frac{\unit[10^{45}]{g\,cm^2}}{I}\right)
  \left(\frac{r}{\unit[10]{kpc}}\right)
  \left(\frac{\unit[600]{Hz}}{f}\right)^2 ,
  \end{split}
\end{equation}
where $\beta$ is an orientation factor, $G$ is the gravitational constant, $r$ is the distance to the source, and $I$ is the moment of inertia.
(One arrives at a similar rough estimate by scaling the upper limits from targeted radiometer analyses with initial LIGO data~\cite{sph_results} to Advanced LIGO sensitivity.)
This is an order of magnitude above the expected strain at $f=\unit[600]{Hz}$ from Scorpius X-1 expected from accretion-torque balance~\cite{sideband_method}, suggesting that a signal might be detectable from another somewhat closer neutron star---especially in light of the fact that additional optimizations are possible, and only a hint of a signal is needed to trigger a coherent matched filtering search, which can boost the significance.

\section{Acknowledgements}
VM's work was supported by NSF grant PHY1204944.
NC's work was supported by NSF grant PHY-1204371.
This is LIGO document P1500015.
\bibliography{asaf}

\end{document}